\documentclass[aps,twocolumn,prl]{revtex4}
\usepackage{amssymb,amsbsy,graphicx,subfigure,bbm,times}
\usepackage{hyperref}
\usepackage[latin1]{inputenc}
\usepackage{amsmath,amsfonts,amssymb,graphics,graphicx,epsfig,color}
\usepackage{subfigure}
\usepackage{fancyhdr,makeidx}
\usepackage{color}

\renewcommand{\prl}{{\it Phys. Rev. Lett.} }

\newcommand{\N}{\cal N}

\newcommand{\id}{\mathbbm{1}}

\newcommand{\tr}{{\rm Tr}\,}
\renewcommand{\det}{{\rm Det}\,}
\newcommand{\gr}[1]{\boldsymbol{#1}}
\newcommand{\be}{\begin{equation}}
\newcommand{\ee}{\end{equation}}
\newcommand{\bea}{\begin{eqnarray}}
\newcommand{\eea}{\end{eqnarray}}

\newcommand{\sig}{\gr{\sigma}}

\newcommand{\eq}[1]{Eq.~(\ref{#1})}

\def\etal{{\it et al.\/}}

\begin{document}
\title{Towards linear phononics and nonlocality tests in ion traps}
\author{Alessio Serafini, Alex Retzker, and Martin B.~Plenio}
\affiliation{Institute for Mathematical Sciences, 53 Prince's Gate, Imperial College London, London SW7 2PG, UK\\
and QOLS, Blackett Laboratory, Imperial College London, London SW7 2BW, UK}

\begin{abstract}
We explore the possibility to manipulate `massive', {\em i.e.}~motional,
degrees of freedom of trapped ions. In particular, we demonstrate
that, if local control of the trapping frequencies is achieved, one
can reproduce the full toolbox of linear optics on radial modes.
Furthermore, assuming only global control of the trapping potential,
we show that unprecedented degrees of continuous variable entanglement
can be obtained and that nonlocality tests with massive degrees of
freedom can be carried out.
\end{abstract}
\maketitle

The last decade saw a boom in the development of experimental
capabilities available for quantum information processing. The
ability to manipulate the information of discrete variables encoded
in polarisation, spin and atomic degrees of freedom has by now
reached very high standards. On the other hand, the control of
continuous variable degrees of freedom is still almost exclusive to
light fields in quantum optics. Even though quantum optical systems
rely on well established tools and are very promising for,
communication tasks, they also suffer from significant drawbacks.
Notably, the entanglement generation in such systems is strongly
limited by the efficiency of parametric processes in nonlinear
crystals; moreover, `static' optical degrees of freedom -- {\em
i.e.}~light resonating in cavities -- are seriously affected by
losses and decoherence over the typical dynamical time scales.

Here, we discuss in detail the possibility of controlling the
radial motion of trapped ions, described by
continuous variable quantum degrees of freedom
which we will refer to as {\em radial modes} \cite{zho2006,nonlinear}.
We highlight the remarkable potential of such modes in
view of the refined technology that has been developed in ion
traps.
In this experimental setting, we demonstrate that 
any linear optical operation can be obtained for radial modes
 of trapped ions by controlling the
individual radial trapping frequencies.
Furthermore we show that,
even if only global control of the trapping potential is possible,
such systems would outperform optical modes in both achievable
degrees of entanglement and decoherence rates. Finally, as an
application, we consider the violation of non locality tests
with radial modes, and show it to be achievable with current
technology, demonstrating the potential of such setups not
only for information processing but also as probes of fundamental
physics.

\noindent{\bf \em The trap} -- We shall consider the radial modes
of $n$ ions of mass $m$ and charge $ze$ in a linear Paul trap
\cite{james98}. Let $\hat{X}_{j}$ and $\hat{P}_{j}$ be the
position and momentum operators associated to the radial degree
of freedom of the $j$-th ion, which is trapped in the radial
direction with angular frequency $\omega_{j}$.
In the following, the {\em longitudinal}
trapping frequency $\nu_t$ will be the unit of frequency and will
set the unit of length as well (equal to
$\sqrt[3]{z^2e^2/(4\pi\varepsilon_{0}m\nu_{t}^2)}$, where
$\varepsilon_0$ is the dielectric constant); also, we shall set
$\hbar=1$. The Coulomb interaction affects the local radial
oscillation frequencies: for convenience, let us then define the
`effective' local radial frequencies $\nu_{j} \equiv
\sqrt{\omega^2_j-\sum_{l\neq j}1/|u_{j}-u_{l}|^{3}}$, $\{u_j\}$
being the equilibrium positions of the ions in the length unit set
by the longitudinal frequency \cite{james98}. Rescaling the canonical
operators according to $\hat{x}_{j}\equiv \sqrt{m\nu_j}\hat{X}_{j}$,
$\hat{p}_{j}\equiv \hat{P}_j/\sqrt{m\nu_j}$, and grouping them
together in a vector of operators
$\hat{R}=(\hat{x}_1,\ldots,\hat{x}_{n},
\hat{p}_j,\ldots,\hat{p}_{n})^{\sf T}$, allows one to express the
global Hamiltonian of the system in the harmonic approximation as
the following quadratic form \be \hat{H} = \frac12 \hat{R}^{\sf T}
\left(\begin{array}{cc}
\gr{\kappa} & 0\\
0 & \gr{\nu}
\end{array}\right)
\hat{R} \label{hami} \; , \ee
where $\gr{\nu}$ is a diagonal matrix:
$\gr{\nu}={\rm diag}\,(\nu_1,\ldots,\nu_{n})$,
while the potential matrix $\gr{\kappa}$ has diagonal entries $\kappa_{jj}=\nu_j$
and off-diagonal entries $\kappa_{jk}=1/(\sqrt{\nu_{j}\nu_{k}}|u_j-u_k|^3)$ for $j\neq k$.
Let us also recall that the
canonical commutation relations can be expressed as
$[\hat{R}_j,\hat{R}_k]=i\Omega_{jk}$, where the $2n\times2n$ matrix
$\Omega$ has entries $\Omega_{j,k}=\delta_{n,k-j}-\delta_{n,j-k}$
for $1\le j,k\le2n$, 
and that, in quantum optics, 
Gaussian states are defined as states with Gaussian characteristic
function: a Gaussian state $\varrho$ is completely determined
by the ``covariance matrix'' (CM) ${\gr\sigma}$, with entries
${\sigma}_{jk} \equiv \tr{[\{\hat R_j , \hat R_k\} \varrho]}/2
-\tr{[\hat R_j \varrho]}\tr{[\hat R_k \varrho]}$ and by the vector
of first moments $R$, with components $R_{j}\equiv \tr{[\hat{R}_j \varrho]}$,
in terms of the vector of canonical operators $\hat{R}$ \cite{rev,martinshash}.

\noindent{\bf \em Linear phononics} -- In the first part of the paper,
we shall assume that the trapping frequencies $\{\omega_j\}$, and
thus $\{\nu_j\}$, can be controlled locally and changed suddenly.
This may be achieved by building small, local radial electrodes, by
adding local optical standing waves \cite{standing}, or in Penning
trap arrays \cite{stahl2005,Ciaramicoli2005}. Our first aim here is
to show how, in principle, this control allows one to perform any
arbitrary `linear optical' operation on the radial modes of the
ions, that is any unitary operation under which, in the Heisenberg
picture, the vector of operators $\hat{R}$ transforms linearly:
$\hat{R}\mapsto S \hat{R}$. The matrix $S$ has to be `symplectic',
{\em i.e.}~$S^{\sf T}\Omega S=\Omega$, to preserve the canonical
commutation relations. Any symplectic operation $S$ on a system of
many canonical degrees of freedom (``modes'') can be decomposed into
a combination of generic single-mode symplectic transformations and
two-mode rotations (``beam splitters'', in the quantum optical
terminology) \cite{pramana,reck94}. It is therefore sufficient for
us to establish the possibility of performing these subclasses of
operations on our system of $n$ ions by manipulating the local
frequencies. {\em Single qubit operations --} In what follows we
assume that the original frequencies of the ions are different but
commensurate, as given by, say, $\nu_{j} = j \nu$, and that $\nu$ is
large enough so that interaction between ions suppressed
\cite{couplenote}.
%
Let us then consider the reaction of the system if the frequency
of the $j$-th ion changes suddenly from $\nu_{j}$ to
$\alpha_{j}\nu_{j}$, for some real $\alpha_{j}$.
The Heisenberg equation of motion for $\hat{x}_j$ and $\hat{p}_j$
can be immediately integrated in such a case, resulting into a
symplectic transformation $S_j(t)$
\be S_j(t) = \left(\begin{array}{cc}
\alpha_j^{\frac12} & 0 \\
0 & \alpha_j^{-\frac12}
\end{array}\right)
\left(\begin{array}{cc}
c & s \\
-s & c
\end{array}\right)
\left(\begin{array}{cc}
\alpha_j^{-\frac12} & 0 \\
0 & \alpha_j^{\frac12}
\end{array}\right)
\label{simple} \, ,
\ee
with $c\equiv\cos(\nu_j\alpha_j t)$ and $s\equiv\sin(\nu_j\alpha_j t)$.
The first and last factor of this decomposition are
`squeezing' operations in the quantum optical terminology,
whereas the second factor is known as a `phase shift'
({\em i.e.}, a rotation in the single-mode phase space).
Combinations of squeezings and phase-shifts make up any
possible single-mode symplectic operation: we thus need
to show that such operations can be implemented individually
on any ion of the system in a controllable manner.

{\em Phase-shift --}
To realise a phase-shift operation on the $k$-th ion,
it is sufficient to change
the frequencies of all the other ions in the same way,
such that $\alpha_k=1$ and $\alpha_j=\alpha\neq 1$ for
$j\neq k$. As apparent from \eq{simple}, after a time
$t_{\alpha}=2\pi/(\nu\alpha)$ one has $S_{j}=\id_{2}$
for $j\neq k$
(let us recall that $\nu_j=j\nu$ by assumption), whereas
the oscillation of the $k$-th ion will have acquired a
phase $\varphi_k=2\pi k/\alpha$ (with no squeezing,
as $\alpha_k$ is kept equal to $1$). If the frequencies
are switched back to the original values after a time
$t_{\alpha}$, the net effect of the evolution is then
analogous to an `optical' phase-shift on the ion $k$.

{\em Squeezing --}
In order to squeeze the state of ion $k$, one can conversely
change only the pertinent frequency, so that $\alpha_k\neq 1$
and $\alpha_j=1$ for $j\neq k$. Then, after a time period
$t=2\pi/\nu$, all the other ions will have returned to the
initial state, while ion $k$ will be squeezed and phase-shifted
according to \eq{simple}. Notice that the phase-shift can always
be corrected by applying the strategy described above.
Let us remark that the degree of squeezing achieved in 
\eq{simple} depends crucially on the phase-shift operation, as
the two squeezing operations act along orthogonal directions and are
the inverse of each other. In the case $\alpha_{k}=(\frac14 +
h)/k$ for $h\in {\mathbbm{N}}$,
the phase-shift can be balanced by a counter-rotation of $\pi/4$ in
phase space and the final squeezing operation is a diagonal matrix
given by ${\rm diag}(\alpha_k,\alpha_k^{-1})$ \cite{limits}. Also
notice that, by placing the ions inside cavities, the squeezing of
the massive degrees of freedom could be transferred to light, so that
radial modes could act as an effective source of squeezing (and
potentially even entanglement) for optical systems as
well.

\noindent{\em Beam-splitters --} Let us now turn to `beam-splitting'
operations(eq.\ref{simple}) between any two radial modes. To this
aim, it is sufficient to bring two modes (hereafter labeled by $j$
and $k$) to the same frequency $\nu=\nu_j=\nu_k$, so that the
Coulomb interaction between them is no longer suppressed.
Switching to an interaction picture, one has the following interaction
Hamiltonian between the two modes: $
\kappa_{jk}\hat{x}_j(t)\hat{x}_k(t) = \kappa_{jk} (a_{j}\,{\rm
e}^{-i\nu t}+a_{j}^{\dag}\,{\rm e}^{i\nu t}) (a_{k}\,{\rm e}^{-i\nu
t}+a_{k}^{\dag}\,{\rm e}^{i\nu t})$, where the ladder operators are
defined as $\hat{x}_j=(a_j+a^{\dag}_j)$. If the frequency $\nu$ is
sufficiently large the rotating wave approximation applies to yield
$ \kappa_{jk} (a_{j}a_{k}^{\dag}+a_{j}^{\dag}a_{k}) \; . $ This
Hamiltonian realises exactly the desired beam splitter-like
evolution, resulting into a symplectic transformation which mixes
$\hat{x}_j$ with $\hat{x}_k$ and $\hat{p}_j$ with $\hat{p}_k$
(rotating such pairs equally, by the angle $\kappa_{jk}t$). For
instance, a `$50:50$' beam splitter is achieved after a time
$t=\pi/(4\kappa_{jk})$. Since the interaction requires a change of
the local frequencies it includes automatically in it a local
operation, which may however be corrected before or after the
`beam-splitting' procedure.

Summing up, we have shown that any symplectic
({\em i.e.} ``linear optical'') operation, including
squeezing, can be implemented for radial
modes of trapped ions by a proper tuning of the
frequencies of the microtraps. {\em Displacement}
operations on individual ions, which shift the
operators $\hat{R}_j$ by a real number, can also be
implemented in microtraps by shifting the radial equilibrium
position of the ion, or as in \cite{Poyatos1996a}.
Because the free evolution rotates the state of
the radial modes in phase space (see \eq{simple} for
$\alpha_j=1$), if the operation is carried out at the proper time
such a shift can be implemented in any direction of phase space and
not only in the positions $\hat{x}_{j}$. The unitary
operator displacing the canonical operators of mode $j$ by,
respectively, $x_{j}$ and $p_{j}$ will be denoted by
$\hat{D}_j(x_j,p_j)$.

These findings show that all the developments based
on {\em Gaussian states} in the quantum optical scenario,
in particular concerning entanglement manipulation
\cite{rev} and information protocols \cite{braunstein05},
could be carried over to radial modes of ion traps
if local control is achieved. In fact, linear optical
operations, complemented by displacements, correspond
to all the unitary transformations that preserve the
Gaussian character of the initial state. 

Notably, even {\em non Gaussian} states can be engineered
in this setup with relative ease, either by entering the
nonlinear regime of the Coulomb interactions or by
exploiting the internal degrees of freedom of the ions.
The latter also allows for Gaussian and non Gaussian
measurements on individual ions: the tomography of
trapped ions, corresponding to {\em homodyning}, was
proposed in \cite{Vogel1995,Poyatos1996a,Bardroff1996} and
partially realized in \cite{Meekhof1996}, while local number
states and parity could be measured using the scheme
suggested and realized for cavity QED in \cite{Gleyzes2007}.
Quite remarkably, such a scheme would allow one to measure
parity on a single copy of the state and run of the apparatus.
In the remainder of the paper, in order to demonstrate the
potential of Gaussian states of radial modes in experimentally
accessible settings, we will consider entanglement generation
and nonlocality tests {\em requiring only global control of
the trapping potential}.

\noindent{\bf \em Entanglement generation} -- The specific Gaussian
situation we shall address starts off from the ground state
$\varrho_{g}$ of Hamiltonian (\ref{hami}) -- with all
frequencies being equal, {\em i.e.}~$\omega_{j}=\omega_i$
for $1\le j\le n$ -- as the initial state (which can be well
approximated in the laboratory by cooling the system to its
ground state \cite{cooling}). Next, the frequency is changed
to $\omega_f$, so that the state $\varrho_{g}$ will not be
stationary anymore under the modified Hamiltonian. For large
$\omega_i$, the initial state $\varrho$ contains very little
entanglement but entanglement builds up during the subsequent
evolution (see \cite{nanos} for an analogous scheme in chains
of nanomechanical oscillators). Entanglement may be quantified by the
logarithmic negativity $E_{\N}\equiv \log_{2}\|\tilde{\varrho}\|_1$,
where $\|\tilde{\varrho}\|_1$ stands for the trace norm of the
`partially transposed' density matrix of the considered system (with
this definition, a Bell pair has $E_{\N}=1$), which is computable
for Gaussian states \cite{gaussneg}.
The ground state of Hamiltonian
(\ref{hami}) is just a Gaussian state with a block diagonal CM
$\sig_{g}=(\sig_{x}\oplus\sig_{x}^{-1})/2$, where
$\sig_{x}=\gr{\nu}^{1/2}(\gr{\nu}^{1/2}\gr{\kappa}\gr{\nu}^{1/2})^{-1/2}\gr{\nu}^{1/2}$,
and vanishing first moments. After the change of potential,
resulting into the new quadratic Hamiltonian $\hat{H}'=\hat{R}^{\sf
T}H'\hat{R}$ (where the `Hamiltonian matrix' $H'$ is implicitly
defined), the evolved state after a time $t$ is a Gaussian state
with CM given by $\sig_{t} = S_{t}\sig_{g}S_t^{\sf T}$, for $S_{t}\equiv\exp(\Omega H't)$.
Furthermore, we have taken into account decoherence
in an environment of phonons with temperature $T$ and `loss rate'
$\gamma$, as the master equation under such conditions admits
Gaussian solutions with CM
${\rm e}^{-\gamma t}\sig_{t}+(1-{\rm e}^{-\gamma t})S_t
\sig_{\infty} S_t^{\sf T}$,
where $\sig_{\infty}\equiv{\rm diag}(\frac12+N_j,\frac12+N_j)$
and $N_j= 1/({\rm e}^{\nu_j/T}-1)$ is the number of thermal
phonons at frequency
$\nu_j$ (setting $k_{B}=1$) \cite{serafozzi05}.

\begin{figure}[t!]
\hspace*{-.6cm}\includegraphics[width=9cm,height=5cm]{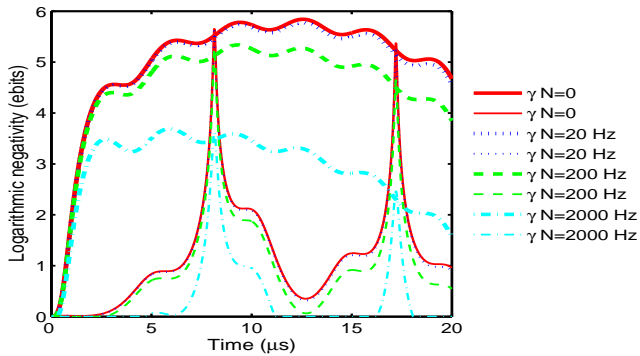}
\caption{Entanglement between first and last ions of the chain 
as a function of time for initial frequency $\omega_i=100{\rm MHz}$,
final frequency $\omega_f=2 {\rm MHz}$, $\nu_t=1\,{\rm MHz}$,
temperature $T=294^{\circ}K$ (corresponding to $N=2\times10^{7}$ thermal phonons)
and different couplings to the environment $\gamma$;
thicker curves refer to ions $1$ and $2$ for $n=2$, 
while thinner curves refer to ions $1$ and $3$ for $n=3$.
The curves for $\gamma=0$ and
$\gamma=10^{-6} {\rm Hz}$ are very close: 
decoherence is almost negligible for such heating rates.
\label{enta}}
\end{figure}

Fig.~\ref{enta} shows that robust entanglement, up to about
$6 \,{\rm ebits}$ of logarithmic negativity, between
two ions can be created in such a setup.
A complete analysis of entanglement and
information propagation through chains of ions will be detailed in
\cite{noialtri}. Here, let us just point out that such degrees of
entanglement are by far out of experimental reach for quantum optics
(where, to the best of our knowledge, $E_{\N}\simeq 1.6$ is the
maximum value so far reported after state reconstruction
\cite{french}). The coupling to the bath of $\gamma=10^{-6} {\rm
Hz}$ (best value considered in the plot) is realistic in view of the
recently observed heating rates in ion traps \cite{Garg,Deslauriers}.
The plot shows that this estimate is extremely encouraging,
especially if compared to the state of the art for quantum optical
cavities and resonators, where loss rates still significantly limit
performances. 
Multipartite entanglement, as well as entanglement between
non-neighboring ions, can also be created with
global control as, when the frequency is switched from $\omega_i$ to
$\omega_f$, all the $n$ ions in the trap start interacting with each
other. For instance, for three ions, $\omega_i=20\,{\rm MHz}$ and
$\omega_{f}=2\,{\rm MHz}$, the initial completely separable state
evolves into a `fully inseparable' Gaussian state (inseparable under
any bipartition of the modes); let us denote the CM of this
state, after an evolution time $t=5\nu_{t}^{-1}=5\mu{\rm s}$, by
$\sig_3$.

\begin{figure}[t!]
\includegraphics[scale=0.8]{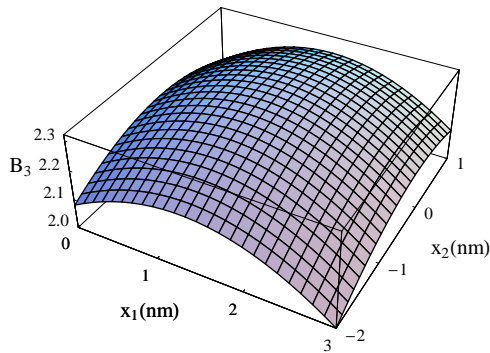}
\caption{Function $B_3$ for $p_1=p_2=p_3=p'_1=p'_3=0$, $x'_1=x'_3=-
x'_2=-3 {\rm nm}$, $p'_2=1 {\rm nm}$,
$0 {\rm nm}\le x_1 \le 3 {\rm nm}$ and $-2 {\rm nm}\le x_2 \le 1 {\rm nm}$
for the Gaussian state with CM $\sig_{3}$ defined in the main text.
Dimensions were reintroduced assuming
$\nu_{t}=1{\rm MHz}$ and Ca ions.
The inequality is violated in the whole displayed region,
where the function reaches a maximum $\simeq 2.32$.}\label{viola3}
\end{figure}

\noindent{\bf\em Nonlocality test} -- Such multipartite entanglement
can be put to use to test quantum nonlocality with massive
particles. Taking advantage of the possibility of performing parity
measurements (in a single shot) and displacements, we will analyse
the violation of the Bell-Klyshko inequalities \cite{klyshko} on the
three ions state with CM $\sig_3$, by the {\em displaced parity}
test as introduced in \cite{banaszek98}. In this instance, the
family of (non-Gaussian) local, bounded, dichotomic observables is
given by
$\Pi_j(x_{j},p_{j})\equiv\hat{D}_j(x_j,p_j)^{\dag}(-1)^{\hat{n}_{j}}\hat{D}_{j}(x_j,p_j)$,
where $\hat{D}_{j}$ and $\hat{n}_j$ are the displacement and number
of phonons operators of ion $j$. The three observers, pertaining to
the three ions, randomly apply two different displacements
[$\hat{D}_j(x_j,p_j)$ and $\hat{D}_j(x'_j,p'_j)$] on their ions and
then measure parity locally.
The expectation value of the operator
$\Pi(R)\equiv\Pi_1(x_{1},p_1)\otimes\Pi_2(x_{2},p_{2})\otimes\Pi_3(x_{3},p_{3})$
is proportional to the Wigner function $W(R)$ of the
composite system evaluated in the point $R=(x_1,x_2,x_3,p_1,p_2,p_3)^{\sf T}$:
$\langle\Pi(R)\rangle=(2/\pi)^3 W(R)$ \cite{wignerparity}.
Such a function is immediately determined for the Gaussian state under consideration:
\be
W(R) = \frac{{\rm e}^{-\frac12R^{\sf T}\sig_{3}^{-1}R}}{\pi^3\sqrt{\det{\sig_{3}}}} \; .
\label{wigner}
\ee
The Bell-Klyshko inequality finally reads:
$
B_3\equiv\frac{8}{\pi^3}|W(x_1,x_2,x'_3,p_1,p_2,p'_3) + W(x_1,x'_2,x_3,p_1,p'_2,p_3)
+ W(x'_1,x_2,x_3,p'_1,p_2,p_3) - W(x'_1,x'_2,x'_3,p'_1,p'_2,p'_3)|\le 2 .
$
Quantum mechanics allows for $2\le B_3\le4$.
As epitomised by Fig.~\ref{viola3}, regions in the space of displacements where the violation
of the inequality is substantial and remarkably stable can be found.
Therefore, this preliminary study opens up very promising perspectives concerning
the violation of Bell inequalities with massive degrees of freedom,
which would be a major, not yet probed, testing ground for fundamental quantum mechanics \cite{retzker05}.
A detailed analysis of the impact of imperfections and noise on such nonlocality tests
will be presented in \cite{noialtri}.

\noindent{\bf\em Conclusions} -- We have demonstrated how the local control of
the trapping frequencies would allow one to reproduce any linear
optical manipulation on radial modes of trapped ions.
We also indicated that phonon detection and homodyne detection  as well
as the implementation of non-Gaussian operations is possible in this setting.
Next, we have
emphasized that, even restricting to global control, such
manipulations enjoy a high efficiency in entanglement generation and
low decoherence rates, along with the possibility of implementing
number and parity measurements with current technology. The
experimental pursuit of the programme outlined in this paper thus
holds considerable promise, concerning both technological
developments, such as the storage and manipulation of quantum
information, and fundamental physical aspects, as in the nonlocality
test for massive degrees of freedom here discussed.

\noindent We thank K.~Pregnell, F.G.S.L.~Brand\~ao, D.M.~Segal,
R.C.~Thompson and T.~Coudreau for helpful discussions. 
This work has been supported
by the European Commission under the Integrated Project QAP, 
by the Royal Society and is part of the EPSRC QIP-IRC.
A.~S.~was funded by a Marie Curie Fellowship.


\begin{thebibliography}{99}
\bibitem{standing} Currently, standing waves realize, at most, trapping potentials
of about $1\,{\rm MHz}$ in experiments where the internal degrees of freedom are controlled.
However, the manipulation of radial modes can tolerate
much higher scattering rates than that of internal degrees of freedom,
and would thus allow for much higher trapping frequencies.
\bibitem{couplenote} The linear scaling of the coupling is not a stringent requirement.
Because the coupling between ion $j$ and $k$ falls off like $|j-k|^{-3}$, one can safely assume,
say, a `triangle' profile (alternately increasing and decreasing) for the couplings on long chains \cite{noialtri}.
Note also that, for a linear scaling, the population of the levels of ion $j$ due to the interaction with $k$ roughly scales as
$1/(1+(\nu/\nu_t)^2|j-k|^5)$: a frequency step $\nu\simeq 20 \nu_t$ is already enough
to make the effects of all the interactions essentially negligible.
\bibitem{limits}In principle, the value of $\alpha_{k}$ is only
limited by
the breakdown of the harmonic approximation \cite{nonlinear},
occurring when the squeezing is comparable to the
ratio between the size of the wave packet and the distance between the atoms. In actual experiments,
the considered setup would thus allow for
$\sqrt{\alpha_{k}} \ll 500$.
Notice that such values are by far out of reach in optical systems,
where squeezings corresponding to $\alpha_k\simeq 8$ were recently reported \cite{takeno07}.
\end{thebibliography}
\end{document}